# Excitation of plasmonic nanoantennas with nonresonant and resonant electron tunnelling

Alexander V. Uskov,*[1,2,3] Jacob B. Khurgin,[4] Igor E. Protsenko,[1,3,5] Igor V. Smetanin,[1,5] Alexandre Bouhelier[6]

[1] P. N. Lebedev Physical Institute, Russian Academy of Sciences, Leninsky pr. 53, 119991 Moscow, Russia; E-mail: alexusk@lebedev.ru
[2] ITMO University, Kronverksky pr. 49, St. Petersburg, 197101, Russia
[3] Advanced Energy Technologies Ltd, Skolkovo, Novaya Ul. 100, 143025 Moscow Region, Russia
[4] Department of Electrical and Computer Engineering, Johns Hopkins University, Baltimore, Maryland 21218, USA
[5] National research nuclear university "MEPhI", Kashirskoe shosse 31, 115409, Moscow, Russia
[6] Laboratoire Interdisciplinaire Carnot de Bourgogne UMR 6303, CNRS-Université de Bourgogne Franche-Comté, 21078 Dijon, France

**Abstract.** A rigorous theory of photon emission accompanied inelastic tunnelling inside the gap of plasmonic nanoantennas has been developed. The disappointingly low efficiency of the electrical excitation of surface plasmon polaritons in these structures can be increased by orders of magnitude when a resonant tunnelling structure is incorporated inside the gap. Resonant tunnelling assisted surface plasmon emitter may become a key element in future electrically-driven nanoplasmonic circuits.

## I Introduction

Plasmonic nanoantennas attract much attention due to their ability at enhancing and controlling effectively the spontaneous emission rate of quantum emitters (molecules, quantum dots and so on) [1-4]. This unique asset has already been put into productive use in optically driven nanoantennas[3-4], primarily for sensing applications. But at the same time, the progress in integrated nanoplasmonic circuits has been impeded by the lack of efficient electrically-pumped sub-wavelength sources of light. For example, semiconductor lasers that enable present day photonic integrated topics suffer from high threshold and low efficiency when scaled down the sub-wavelength dimensions [5]. Given that, electrically-driven plasmonic nanoantennas appear to be a logical approach to the problem due to its apparent simplicity, as no nanoscale p-n junction needs to be formed and the light confinement is easily achieved in the vicinity of surface plasmon polariton (SPP) resonance.

Excitation of plasmonic oscillations (both propagating and localized) with electron tunnelling is far from being a new topic – it predates the first appearance of the term "plasmonics" by decades., but as we show below, it deserves a second look. Starting with the pioneering 1976 work [6] by Lambe and McCarthy, tunnelling excitation of surface plasmonic waves in planar Metal-Insulator-Metal (MIM) structures had been an object of plentiful experimental[6-15] and theoretical [16-20] studies. These investigations had been given an impetus by the invention of Scanning Tunnelling Microscope (STM) at the end of 1980's as numerous intensive studies of excitation of plasmonic modes with STM tips has been since performed [21-29]. With the advent of nanophotonics the focus of research on SPP electrical excitation has gradually shifted to the development of miniature light sources as can be learned from recent reports [30-41] and review [42] by García de Abajo.

The results of the experimental measurements often differ from each other and from theoretical estimates by as much as an order of magnitude or more[30-41]. This is expected, given the great variety of experimental conditions as well as equally great diversity of theoretical approaches. Overall, though, it comes as no surprise that the most serious disadvantage of the electron tunnelling mechanism for the electrical excitation of SPP is its low quantum efficiency (QE), typically in the range of $10^{-4}$-$10^{-6}$. The probability of inelastic tunnelling accompanied by the emission of a quantum of electromagnetic field competes with the far stronger elastic electron tunnelling in which no electromagnetic radiation is emitted. The latter process is very efficient (it is this process, after all, that is exploited in STM), and occurs on a femtosecond scale. The electron-photon coupling is typically weak, so the electrons tunnel through the gap before they can effectively interact with the electromagnetic field. Even though the density of the electromagnetic energy is enhanced by the SPP, the resulting Purcell enhancement is not sufficient to raise the QE of SPP emission to more than $10^{-4}$ [36].

Based on these simple considerations, one way to increase the QE of SPP emission would be to "delay" the elastically tunnelling electron inside the gap by creating additional barriers to it, such as multilayer metal-dielectric structures with tunnelling barriers containing quantum wells (QW) [20]. As the resonantly tunnelling electron remains confined inside the QW for a prolonged period of time, the probability of making a downward transition and emitting an SPP in the process is increased. In this letter we subject this intuitive idea to a rigorous theoretical test by comparing the QE of SPP excitation in nanoantennas with and without resonant tunnelling and conclude that indeed the QE in the resonant tunnelling structure can be raised by several orders of magnitude to the levels where it may become practical to use plasmonic nano-antennas as the source of choice in future nanophotonics circuits.

## II Theoretical Formalism

The key to obtain a strong emission in a nanostructure is to take advantage of the Purcell effect. This requires the existence of an electromagnetic mode whose effective volume is substantially smaller than $\lambda_o^3$ where $\lambda_o$ is the wavelength in vacuum. Various SPP modes in diverse metallic nanostructures satisfy this condition, for instance propagating slab and gap SPPs as well as localized SPPs in nanoparticles, dimers, bow-tie antennas and so on [2-4]. Having a small effective volume, however, is not sufficient for high QE of radiation. Small-volume tightly confined modes are only weakly coupled to free-space and instead of being radiated, the energy of these SPP modes is dissipated in the metal. For this reason, the external (or radiative) QE of various nanoplasmonic structures coupled to quantum emitters had been optimized with extensive numerical calculations, and a large body of literature is available. In this work we concentrate only on the "internal" QE that is independent on the coupling between the SPP and vacuum radiation. In the following, we adapt a rather simple model shown in Fig.1 and consisting of a metal nanowire of radius $a$ and length $L$ buried inside a dielectric medium with the dielectric permittivity $\varepsilon_d$. The nanowire is cut along its length, and a thin dielectric layer (or several layers of various dielectrics (semiconductors) and metals) is inserted between the halves of the nanowire. These two halves are the electrodes and may be electrically contacted by an external power source. Since our goal is to compare the QE of various tunnelling structures in relative terms, the relation between their QE will hold for any SPP mode. In order to enhance the absolute external QE one only needs to follow the well-established studies available in the literature [2,4].

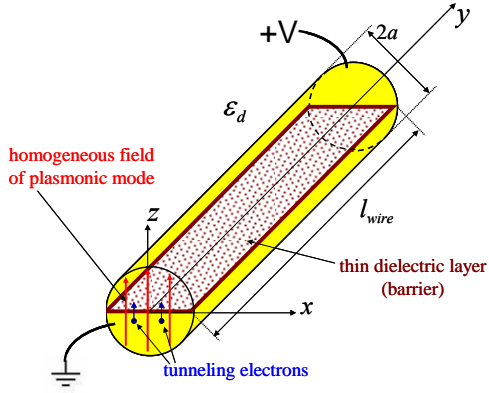

**Fig. 1** An electrically-contacted tunnelling plasmonic nanoantenna: a metal nanowire of length $l_{wire}$ and of radius $a$ is cut along the *y*-axis; a thin dielectric barrier is inserted between the two halves of the nanoantenna. Electrons (circles) tunnel when the voltage $V$ is applied between the two halves. Red arrows along *z*-axis indicate the homogeneous electric field of the plasmonic mode.

The presence of a dielectric layer creates a potential barrier $U_b(z)$ between the electrodes as shown in the energy diagram in Fig.2. When the external voltage $V$ is applied between these electrodes, the tunnelling current flows. Electrons may tunnel either elastically, i.e., without loss of energy during the tunnelling (green arrow), or inelastically, when a fraction of the energy $\hbar\omega_{sp}$ is used to excite an SPP in the nanoantenna (red arrows in Fig.2). The QE of nanoantenna excitation is then defined as $\eta = I_{inel}/(I_{inel} + I_{el})$ where $I_{el}$ and $I_{inel}$ are the tunnelling currents due to elastic and inelastic tunnelling, respectively, see

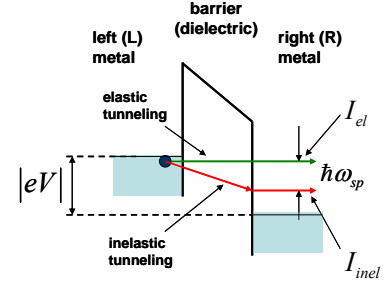

**Fig. 2** Energy diagram of the MIM structure: $|eV|$ is the difference between the Fermi-levels of left (L) and right (R) electrodes. Electrons may tunnel between the electrodes elastically ($I_{el}$, green arrow) or inelastically ($I_{inel}$, red arrows) with emission of an SPP with energy $\hbar\omega_{sp}$.

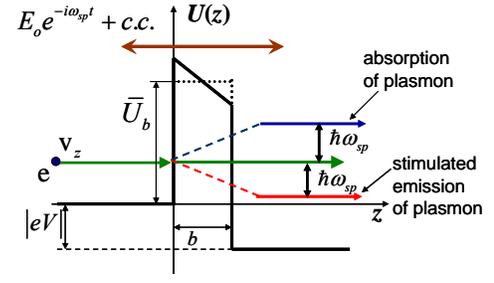

**Fig. 3** Stimulated emission and absorption of a SPP with an energy $\hbar\omega_{sp}$ in the presence of a homogeneous SPP field $E(t) = E_o \cdot e^{-i\omega_{sp}t} + c.c.$ (brown arrow). $U(z)$ is the electron potential. An electron impinges the barrier from the left with the velocity $v_z$. During tunnelling the electron may absorb (blue arrow) or emit (red arrow) an SPP. The probabilities of both the emission and absorption are proportional to the square of the SPP field amplitude $|E_o|^2$. Dotted line shows the averaged potential $\overline{U}_b$ in the barrier used in the calculations.

Inelastic tunnelling with the excitation of a SPP is best described as the spontaneous emission of the SPP by an electron during its transition from upper energy level in left electrode to lower level in the right electrode (see Fig. 2). To find the probability of this spontaneous emission, we use Einstein's relations between the probabilities of stimulated and spontaneous emissions. The probability of stimulated emission (see Fig.3) of the quantum of energy $\hbar\omega_{sp}$ in the SPP mode of amplitude $E_o$ is $p^{stim} = C^{stim} \cdot |E_o|^2$, where $C^{stim}$ is a constant proportional to the second Einstein's coefficient. The energy in the SPP mode $W_{mode}$ is proportional to the square of the amplitude, $W_{mode} = c_w \cdot |E_o|^2$, where $c_w$ is the coefficient proportional to the effective volume of the SPP mode. The probability of spontaneous emission $p^{spon}$ can then be readily found by substituting $|E_o|^2 \to |E_{vac}|^2$ where the square of the "vacuum field" $|E_{vac}|^2$ is found from the relationship $W_{mode}^{vac} = c_w \cdot |E_{vac}|^2 = \hbar\omega_{sp}/2$. Under the assumption that spontaneous emission is nothing but a stimulated emission caused by "vacuum fluctuations" of the SPP mode [13, 28], $p^{spon} = (\hbar\omega_{sp}/2) \cdot (C^{stim}/c_w)$.

In our example we consider the excitation of the fundamental SPP mode of the nanowire near the resonant frequency $\omega_{sp} = \omega_p/\sqrt{1+\varepsilon_d}$ where $\omega_p$ is the metal plasma frequency of the metal of nanowire[1]. Inside the nanowire, the electric field of the mode is homogeneous (see Fig.1), $\mathbf{E}_{inside}(t) = \mathbf{E}_o \cdot e^{-i\omega_{sp}t} + c.c.$, and normal to the dielectric layer. The energy of the mode $W_{mode}$ can be calculated as $W_{mode} = 2l_{wire} \cdot \pi a^2 \varepsilon_o (1+\varepsilon_d)|E_o|^2$ where $\varepsilon_o$ is the vacuum permittivity (Electronic Supplementary Information, SI-1).

The probability of stimulated emission can be found by applying Fermi's Golden Rule (see, for instance[43]) to the one dimensional tunnelling problem shown in Fig.3. Assuming that the incident electron travels with velocity $v_z = \hbar k_{zL}/m$ where $k_{zL}$ is the longitudinal wave vector and $m$ is the electron mass this probability is

$$p^{stim}(v_z) = \frac{m}{\hbar^3} \cdot \frac{1}{k_{zR}} \cdot |\langle \psi_L(z)|H'|\psi_R(z)\rangle|^2 =$$
$$= \frac{e^2}{m\omega_{sp}^2} \frac{1}{\hbar k_{zR}} \cdot |\langle \psi_L(z)|\frac{\partial}{\partial z}|\psi_R(z)\rangle|^2 \cdot |E_o|^2 \quad (1)$$

where $\hat{H}' = -e\hbar/(m\omega_{sp}) \cdot E_o^* \cdot \partial/\partial z$ is the interaction Hamiltonian of the electron with the field; $e$ is the electron charge, $\psi_{L,R}(z)$ are the wavefunction in the left and right electrode with the energies $E_L$ and $E_R = E_L - \hbar\omega_{sp}$ respectively, and $k_{zR} = \sqrt{k_{zL}^2 + 2\hbar^{-2}m|eV| - 2m\hbar^{-1}\omega_{sp}}$ is the longitudinal wave vector of the electron in the right electrode. Note that in order to have the dimensionless probability in Eq. (1) the incident wave function $\psi_L(z)$ must be normalized to the unit flux of probability, while the function $\psi_R(z)$ is normalized to the unit density of probability[43].

## III Results and Discussions

We now apply the formalism of Eq. (1) to single and double tunnelling barriers as illustrated in Fig.4 a and b, respectively.

**A Inelastic tunnelling through a single barrier**

For the single barrier (SB) (Fig.4a) the transition matrix element can be evaluated (as shown in Electronic Supplementary Information, SI-2) and the probability for inelastic non-resonant tunnelling in single barrier structure accompanied by stimulated emission of SPP is

$$p_{SB}^{stim}(v_z) \sim e^{-2\kappa_L b} \cdot \frac{e^2}{\hbar^2 \omega_{sp}^4} \cdot v_z^2 \cdot |E_o|^2 \quad (2)$$

where $b$ is the barrier width, $\kappa_L = \sqrt{2m\hbar^{-2}\bar{U}_b - k_{zL}^2}$ is the decay constant of $\psi_L(z)$ of the incident electron inside the barrier and $\bar{U}_b$ is the average height of the barrier. In Eq. (2), we dropped the prefactor which is of the order of unity (the so-called exponential approximation[43] in tunnelling theory). Also we assumed that $e^{-(\kappa_R - \kappa_L)b} \ll 1$, where $\kappa_R = \sqrt{2m\hbar^2(\bar{U}_b + \hbar\omega_{sp}) - k_{zL}^2}$ is the decay constant of the final wave function $\psi_R(z)$. The condition $e^{-(\kappa_R - \kappa_L)b} \ll 1$ can be phenomenologically interpreted as saying that the SPP is emitted when the tunnelling electron collides with the drop of the potential at the interface of the left electrode and the barrier.

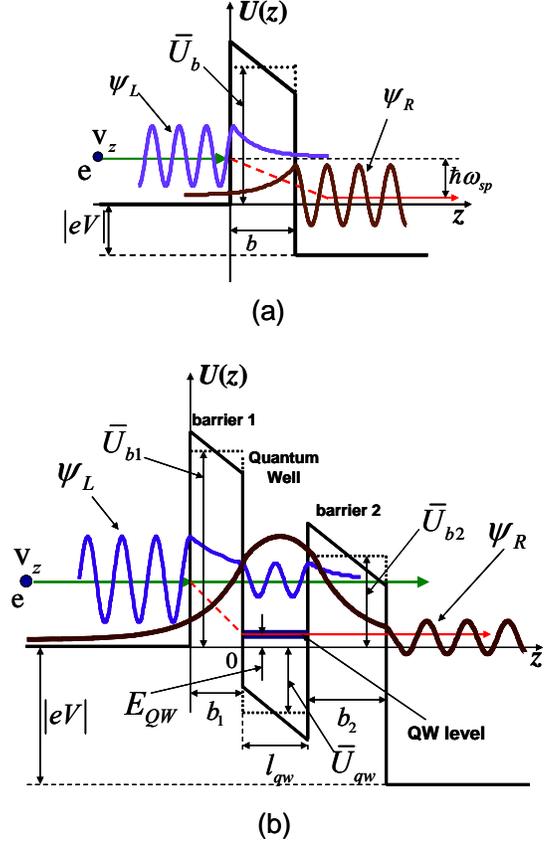

**Fig. 4** (a) Potential barrier $U(z)$ for tunnelling electron in the case of a single dielectric layer between two metal electrodes with a schematic illustration of the wave functions $\psi_L(z)$ and $\psi_R(z)$ of an electron in the left and in the right electrodes, respectively. The tails of these wave functions penetrate and overlap into the barrier. (b) Potential barrier $U(z)$ for a double-barrier structure with a Quantum Well located between two barriers, together with a schematic illustration of the wave functions $\psi_L(z)$ and $\psi_R(z)$ of electron in the left and right metals. The energy level of the QW is shown as dark blue. The green arrow is the elastic tunnelling, the red arrow is the inelastic tunnelling with emission of plasmon and the dotted lines show the averaged potentials insides the barriers and the QW.

Using the aforementioned relations between stimulated and spontaneous emission, the probability of inelastic tunnelling in single barrier structure accompanied with spontaneous emission then becomes

$$p_{SB}^{inelast}(v_z) \sim e^{-2\kappa_L b} \cdot \frac{\alpha_{fs}}{4\pi^2(1+\varepsilon_d)} \cdot \frac{v_z^2}{c^2} \cdot \frac{\lambda_{sp}^3}{V_{nano}} \quad (3)$$

where $\alpha_{fs} = e^2/(4\pi\hbar\varepsilon_o c) = 1/137$ is the fine structure constant; $V_{nano} = l_{wire} \cdot \pi a^2$ is the antenna volume, and $\lambda_{sp} = 2\pi c/\omega_{sp}$. Using the same exponential approximation the probability of elastic tunnelling is

$$p_{SB}^{elast}(v_z) \sim e^{-2\kappa_L b} \quad (4)$$

The current densities for both elastic and inelastic tunnelling are evaluated by performing a summation over the states in the left electrode under the assumption that the voltage is high enough to have the final state in the right electrode empty[44]

$$J_z = e \cdot \frac{m k_B T_e}{2\pi^2 \hbar^3} \cdot \int_0^{+\infty} dE_z \cdot p(E_z) \cdot \ln\left(1 + e^{-\frac{E_z - \varepsilon_F}{k_B T_e}}\right) \quad (5)$$

where $p(E_z)$ is the probability of either elastic or inelastic tunnelling, $E_z = mv_z^2/2$ is the energy of the incident electron; $T_e$ is the electron temperature; $k_B$ is Boltzmann constant, and $\varepsilon_F$ is the Fermi energy of the metal. Substituting Eq. (3) and Eq. (4) into Eq. (5), one can evaluate the current densities $J_{SB}^{el}$ and $J_{SB}^{inel}$ due to, respectively, elastic and inelastic tunnelling in single barrier structure with nonresonant tunnelling, and therefore the QE of nanoantenna excitation

$$\eta_{nonres} = \frac{J_{SB}^{inel}}{J_{SB}^{inel} + J_{SB}^{el}} \approx \frac{\alpha_{fs}}{4\pi^2(1+\varepsilon_d)} \cdot \frac{v_F^2}{c^2} \cdot \frac{\lambda_{sp}^3}{V_{nano}} \quad (6)$$

where $v_F$ is the Fermi velocity. Given the fact that neither the Fermi velocity ($v_F = 1.4 \cdot 10^6$ m/s in gold) nor the dielectric constant ($\varepsilon_d \sim 13$ in Si) vary significantly in different materials, and stressing the fact that the $\eta_{nonres}$ does not depend on the barrier height and width, we can re-write Eq. (6) as $\eta_{nonres} \approx 3\times 10^{-10} \lambda_{sp}^3/V_{nano}$. In other words, the electrical excitation efficiency of the SPP depends only on what can be loosely defined as the Purcell Factor $\lambda_{sp}^3/V_{nano}$ [45], where in gold $\lambda_{sp} = 516$ nm. For a 100nm long nanoantenna of 20nm radius $\lambda_{sp}^3/V_{nano} \sim 10^3$ and the QE of excitation is less than $10^{-6}$. Reducing the nanoantenna dimensions by, say, an order of magnitude $L \sim 10$nm, $a \sim 2$nm would raise the QE to $10^{-3}$. Aside from fabrication difficulties minimizing the dimensions might not be a viable strategy, as the radiative property of the nanoantenna would decrease greatly. In general, as shown in [3, 36] the maximum spontaneous enhancement factor can be optimized, but then radiative efficiency of the antenna decreases to less than 10%, therefore the maximum external QE of the inelastic tunnelling through a single barrier cannot exceed $10^{-4}$ under the best of circumstances.

This disappointing result can ultimately be traced to the weak coupling of an electron with the SPP's electric field. It can be intuitively interpreted in two ways. In a first interpretation the SPP is emitted when the electron collides with the metal surface and that process happens just once. In a second explanation the SPP is emitted only when the tunnelling electron is inside the gap, and that time is very short. Either one of these interpretations immediately leads to the direction which should be pursued in order to increase the QE – one should strive to increase the number of collisions (or equivalently the dwelling time in the gap) by using resonant tunnelling through the MIMIM structures with double barriers.

An alternative way would obviously be to suppress the elastic tunnelling completely by which would require using a highly doped semiconductor as a right electrode and then aligning its bandgap to block the elastic tunnelling. However, given the fabrication difficulties, this path to the QE enhancement seems to be less realistic than the one that follows.

**B Double barrier structure with resonant tunnelling**

As an example of a structure with resonant tunnelling we consider the Double Barrier (DB) structure with a Quantum Well (QW) placed between the barriers. It is a multilayer structure Metal-Insulator-Metal-Insulator-Metal (MIMIM) or Metal-Insulator1-Insulator2-Insulator3-Metal (M-I1-I2-I3-M) – see Fig. 4b. The principal feature of these structures is the presence of a QW inside the potential barriers with a discrete, or quasi-discrete, energy level. This QW with a discrete energy level may lead to resonance tunnelling [46-48]. In Fig. 4b, the left and right barriers have different thicknesses. This asymmetry significantly reduces the *elastic* resonant transmission of the structure from 100% down to a very low value [46] while still ensuring the enhancement of the density of probability for electron inside the well. One should note that one possible way for realizing a resonant tunnelling could be the insertion of quantum dot(s) or molecule(s) inside the nanogap of plasmonic nanoantennas – see, for instance, [49-50].

Figure 4b illustrates the behaviour of the wave functions $\psi_L(z)$ and $\psi_R(z)$ of the final state after the SPP emission – for details, see Electronic Supplementary Information, SI-3. If the energy $E_R$ of the state coincides with the energy $E_{QW}$ of the QW level, we have an exponential increase of the amplitude of $\psi_R(z)$ inside the barrier 2 that leads, in turn, to an increase of the matrix element $\langle \psi_L(z)|\partial/\partial z|\psi_R(z)\rangle$, and a corresponding increase of the probabilities of tunnelling, Eq. (1). This behaviour of $\psi_R(z)$ describes the well-established fact that when the incident electron penetrates into the QW, it populates the energy level during prolonged period of time, just as a photon dwells for a long period of time in a resonant Fabry-Perot cavity. Multiple reflections of the electron from the barriers enhance the coupling of the electron with the electromagnetic field, thereby boosting the probability of SPP emission. Note that the elastic tunnelling in Fig. 4b (green arrow) is nonresonant; hence the QW's discrete level can only enhance the spontaneous emission of SPP. Since it does not enhance the elastic contribution, the QE does gets improved.

When the energy $E_R = E_L - \hbar\omega_{sp}$ of the final state is close to the QW level energy $E_{QW}$ (see SI-3), we can calculate the matrix element in the probability of excitation of plasmonic nanoantenna with resonant tunnelling in DB structure as

$$p_{DB,res}^{inelast}(v_z) \sim$$
$$\sim e^{-2\kappa_{L1}b_1} \cdot \frac{E_{QW} - \bar{U}_{qw}}{2\pi \cdot \Delta E_{QW}} \frac{1}{1+\left(\frac{E_R - E_{QW}}{\Delta E_{QW}}\right)^2} \cdot \frac{\alpha_{fs}}{4\pi^2(1+\varepsilon_d)} \cdot \frac{v_z^2}{c^2} \cdot \frac{\lambda_r^3}{V_{nano}} \quad (7)$$

where $b_1$ ($b_2$) is the thickness of left (right) barrier, $\kappa_{L1} = \sqrt{2m\bar{U}_{b1}/\hbar^2 - k_{zL}^2}$ is the decay constant of the initial wave function $\psi_L(z)$ of the incident electron inside the left barrier; $\bar{U}_{b1}$ ($\bar{U}_{b2}$) is the average height of the left (right) barrier, $\Delta E_{QW} = e^{-2\kappa_{R2}b_2} \cdot (E_{QW} - \bar{U}_{qw})/2\pi$ is the broadening of the level in QW, and $\bar{U}_{qw}$ is the average potential in the QW, $\kappa_{R2} = \sqrt{2m(\bar{U}_{b2} + \hbar\omega_{sp})/\hbar^2 - k_{zL}^2}$ is the decay constant of the final wave function $\psi_R(z)$ of the electron in the right barrier after the emission of the SPP $\hbar\omega_{sp}$. It is clear that the width $\Delta E_{QW}$ of the resonance in Eq. (7) decreases exponentially with increasing barrier width $b_2$ and the barrier height $\bar{U}_{b2}$ with a corresponding increase of the probability $p_{DB,res}^{inelast}(v_z)$ of Eq. (7). This can be related to the decreased transmission of the right barrier ($\sim e^{-2\kappa_{R2}b_2}$) with larger barrier width $b_2$ and $\bar{U}_{b2}$, and to the corresponding increase of the electron dwell time inside the QW. The probability of elastic tunnelling for the considered double barrier structure can be estimated as

$$p_{DB}^{elast} \sim e^{-2\kappa_{L1}b_1} \cdot e^{-2\kappa_{L2}b_2} \qquad (8)$$

where $\kappa_{L2} = \sqrt{2m\bar{U}_{b2}/\hbar^2 - k_{zL}^2}$ is the decay constant of the initial wave function $\psi_L(z)$ of incident electron inside the right barrier.

One can see by comparison of Eqs. (7)-(8) with Eqs. (3)-(4) that the QE can be increased substantially in the vicinity of resonance. Nevertheless, one should stress here that Eq. (7) is obtained under the assumption of a fully coherent transit of electrons through the barriers. In other words, we have neglected completely the possibility for tunnelling electron to experience electron-phonon and electron-electron collisions, which break the coherency and broaden the QW's energy level. One may take into account the broadening with the following semi-phenomenological procedure: in Eq. (7) we substitute

$$\Delta E_{QW} \to \Delta E_{QW} + \hbar/\tau_2 \qquad (9)$$

where $\tau_2$ is the dephasing time of electron due to electron-phonon and electron-electron collisions, which is $\sim(10-100)\,\text{fs}$ depending on the structure materials. Equation (9) imposes that the minimal width of QW level is dictated by the value of $\hbar/\tau_2$.

Consider now the QE of nanoantenna excited by a single electron

$$\eta_{1res} = \frac{p_{DB,res}^{inelast}}{p_{DB,res}^{inelast} + p_{DB}^{elast}} = \frac{\eta_{nonres}}{D_{L2}\cdot\left(D_{R2} + \dfrac{2\pi\hbar/\tau_2}{E_{QW} - \bar{U}_{qw}}\right) + \eta_{nonres}} \qquad (10)$$

where $D_{R2} = e^{-2\kappa_{R2}b_2}$ is the transmission of the right barrier (the index 2) for the electron in its final state (after an SPP emission), $D_{L2} = e^{-2\kappa_{L2}b_2}$ is the transmission of the left barrier (the index 2) for the electron in its initial state (before SPP emission). Formula (10) clearly demonstrates that when it comes to achieving higher QE's, double barrier structures with plasmon-assisted resonant tunnelling hold an undisputable advantage over the conventional single barrier structures with non-resonant tunnelling. On one hand, one sees that the QE may reach values close to unity by decreasing the transmission coefficients $D_{R2}$ and $D_{L2}$ of the right barrier. On the other hand, it shows also the effect of the dephasing time $\tau_2$: if $D_{R2} < (2\pi\hbar/\tau_2)/(E_{QD} - \bar{U}_{QD})$, further decrease of the transmission $D_{R2}$ does not result in increase of the QE. That is, of course, because the decrease of $D_{R2}$ does not lead to a narrowing of the resonance in Eq. (7) since its minimal width is achieved already. So, if $D_{R2} < (2\pi\hbar/\tau_2)/(E_{QW} - \bar{U}_{QW})$ the improvement of $\eta_1^{res}$ occurs only through a decreasing $D_{L2}$. Thus, in order to reach, say $\eta_1^{res} \sim 0.1$, one must have $D_{L2} \sim 10 \cdot \eta_1^{nonres} \cdot (E_{QW} - \bar{U}_{QW})/(2\pi\hbar/\tau_2)$.

Dependence of $\eta_1^{res}$ on the transmission $D_{L2}$ shown in Fig. 5 illustrates the behaviour. In the calculations, $\eta_1^{nonres} = 10^{-6}$, the ratio $\kappa_{L2}/\kappa_{R2}$ is assumed to be constant, and equals to 0.5. The upper curve is for $\tau_2 = \infty$ (i.e., it is assumed that the transit of electron through barrier structure is completely coherent). If this condition, $\eta_1^{res}$ is close to unity for small transmission coefficients. For the cases $\tau_2 = 100\,\text{fs}$ and $\tau_2 = 10\,\text{fs}$, the quantum efficiency $\eta_1^{res}$ reaches values larger than ~0.1 for $D_{L2} \sim 10^{-3} - 10^{-4}$. This takes place through the suppression of elastic tunnelling. Of course, the QE defined here only relates the rates of photon emission accompanied and resonant tunnelling and does not include inelastic nonradiative processes, such as phonon-assisted tunnelling which is bound to limit the QE. Furthermore, as the QE increases to very high value the absolute value of the tunnelling probability drastically drops down, This is discussed below as we turn our attention to the estimation of the overall tunnelling current.

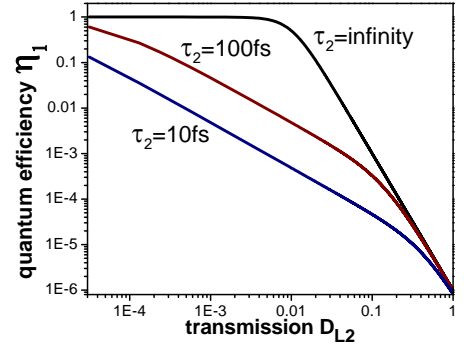

**Fig. 5** Dependence of the quantum efficiency of nanoantenna excitation by a single resonant electron as a function of the transmission coefficient $D_{L2}$ of the second (left) barrier for an electron in its initial state for three values of the dephasing time $\tau_2$: black ($\tau_2 = \infty$), brown ($\tau_2 = 100\,\text{fs}$), dark blue ($\tau_2 = 10\,\text{fs}$).

Having determined the QE of SPP emission for a single electron, one can now sum over all the incoming electrons. Using Eqs. (7) and (8) in Eq. (5) [see Electronic Supplementary Information, SI-4], one find the current densities due to elastic $J_{DB}^{el}$ and inelastic $J_{DB}^{inel}$ tunnelling in DB structure, and then estimate the quantum efficiency $\eta_{res}$ for resonant tunnelling DB structure as

$$\eta_{res} = \frac{J_{DB,res}^{inel}}{J_{DB,res}^{inel} + J_{DB}^{el}} = \frac{\eta_{nonres}}{D_{L2}(\varepsilon_F) \cdot \dfrac{2\varepsilon_F}{E_{QW} - \bar{U}_{qw}}\left(\dfrac{\tilde{\varepsilon}}{\varepsilon_F}\right)^2 + \eta_{nonres}} \qquad (11)$$

where $\tilde{\varepsilon} = \left[\kappa_1^{\varepsilon_F}b_1/(\bar{U}_{b1} - \varepsilon_F) + \kappa_2^{\varepsilon_F}b_2/(\bar{U}_{b2} - \varepsilon_F)\right]^{-1}$ with $\kappa_j^{\varepsilon_F} = \sqrt{2m(\bar{U}_{bj} - \varepsilon_F)/\hbar^2}$ ($j=1,2$). Figure 4 illustrates Eq. (11), and one can see that with a proper design the QE in resonant tunnelling structures can reach values higher than 10%. Of course, while the QE of SPP emission is high for thick barriers, the overall current is very small, and so is the absolute rate of SPP emission. To increase that rate to the point where the amount of radiated light is sufficient one will always be compelled to operate away from the maximum QE point. This optimization needs to be performed concurrently with the optimization of the external efficiency of the nanoantennas and will be the subject of future work.

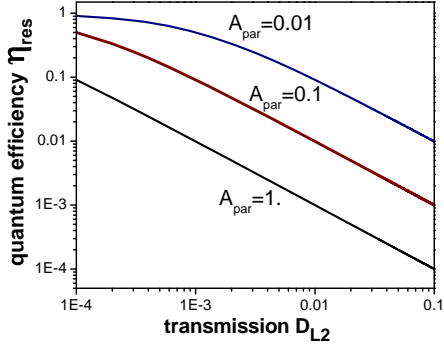

**Fig. 6** Dependence of the quantum efficiency $\eta_{res}$ of nanoantenna excitation for resonant tunnelling structure on the transmission coefficient $D_{L2}$ of the second (left) barrier for electron in its initial state for various values of the parameter $A_{par} = 2\varepsilon_F / (E_{QW} - \bar{U}_{qw}) \cdot (\tilde{\varepsilon}/\varepsilon_F)^2$: black ($A_{par} = 1$), brown ($A_{par} = 0.1$), dark blue ($A_{par} = 0.01$); the quantum efficiency for nonresonant tunnelling structure $\eta_{nonres} = 10^{-5}$.

## VI Conclusions

In conclusion we have developed a rigorous theory of all-electrical excitation of plasmonic nanoantennas by means of photon emission accompanied tunnelling. The quantum efficiency of plasmon excitation is disappointingly low as the emission process competes with the much more favourable elastic tunnelling contribution. This efficiency, however, can be increased manifold if the tunnelling electron can be slowed down in a resonant structure incorporating a quantum well thus making electrically driven nanoantenna emitters practical.

## Acknowledgements

This work of AU was financially supported in part by the Government of the Russian Federation (Grant 074-U01) through ITMO Visiting Professorship program. JK would like to thank NSF MIRTHE ERC for the support. AB acknowledges the funding from the European Research Counsel under the European Community's Seventh Framework program FP7/2007-2013 Grant Agreement 306772, the label ACTION (ANR-11-LABX-01-01), and the regional program PARI